Reproducibility of room temperature ferromagnetism in $Zn_{0.95}Mn_{0.05}O$ and its understanding

R.N. Bhowmik[1*], Asok Poddar[2] and A. Saravanan[1]

[1]Department of Physics, Pondicherry University, R. Venkataraman Nagar, Kalapet, Pondicherry-605014, India.

[2]Experimental Condensed Matter Physics, Saha Institute of Nuclear Physics, 1/AF Bidhannagar, Kolkata-700064, India.

[1*]Corresponding author (RNB): rnbhowmik.phy@pondiuni.edu.in, rabindranath.bhowmik@saha.ac.in

**Abstract:**

The present work reproduces the room temperature ferromagnetism by doping magnetic Mn atoms in diamagnetic ZnO. The ferromagnetic ordering is extended up to 640 K in mechanical milled $Zn_{0.95}Mn_{0.05}O$ samples. The bulk and nanocrystalline samples are stabilized in hexagonal crystal structure with space group p63mc. The grain size and lattice strain of the samples, estimated from XRD spectrum using Williamson-Hall plot, showed a significant variation with milling time. Surface structure (morphology, distribution of grains and elements) is observed to be reasonably good from SEM picture and EDX spectrum at room temperature. The ferromagnetic ordering in bulk, milled and alloyed samples is primarily due to the diffusion of $Mn^{2+}$ ions into the lattice sites of ZnO. The enhancement of magnetic moment and ferromagnetic ordering temperature with reducing the grain size is discussed in terms of the existing theoretical predictions and experimental works. The role of mechanical milling induced disorder for the enhancement of room temperature ferromagnetism in $Zn_{0.95}Mn_{0.05}O$ (dilute magnetic semiconductor) is also highlighted.

## 1. Introduction:

Dilute magnetic semiconductors (DMS) are the solid solution of non-magnetic semiconductor doped with magnetic elements (Mn, Fe, Co and Ni). DMS emerges as the candidate for the potential applications in spintronic devices, in which spin of the electron plays a major role in reading and writing the information in recording medium [1, 2]. In order to search the ferromagnetic DMS for room temperature applications, Dietl et al. [3] predicted the Curie temperature ($T_C$) higher than room temperature for 5% Mn doped ZnO, inspite of the fact that ZnO is a well known direct band gap (3.37eV) non-magnetic (diamagnetic) semiconductor. The interest of doping magnetic element in ZnO stems from room temperature ferromagnetism (RTF) [4, 5]. Earlier reports also suggested the multifunctional properties of ZnO based DMS that can be applied in the field of semiconductor devices [6, 7] and acoustic-optical devices [8, 9]. The interesting aspect of studying such materials is to understand whether the ferromagnetism is intrinsic or extrinsic of the doping effect.

The origin of ferromagnetic ordering is still a matter of discussion for Mn doped ZnO (i.e., Mn-ZnO). The reports [10-16] indicated a wide range of magnetic ordering (ferromagnetic, antiferromagnetic, paramagnetic and spin glass) at room temperature (300 K) for Mn doped ZnO. Different views were presented in literature to realize the magnetism of ZnO based DMS. One mechanism is carrier mediated magnetism. Monte Carlo simulation [13] explained RTF in Mn doped ZnO by considering two competing indirect exchange interactions, i.e., antiferromagnetic superexchange and an oscillating carrier mediated interaction. The simulation showed that p-type Mn-ZnO (Mn up to 20%) is paramagnet for hole concentration $1 \times 10^{16}$ to $1 \times 10^{18}$ cm$^{-3}$ and ferromagnet for hole concentration $1 \times 10^{19}$ cm$^{-3}$ and above, as also predicted by Dietl et al. [3] and N.A. Spaldin [17]. M. Ivill et al. [12] verified the enhancement of carrier mediated ferromagnetism due to co-doping effect in ZnO matrix. On the other hand, antiferromagnetism [14] or spin glass [18] or no phase transition [13] was suggested for n-type Mn doped ZnO samples. Next, the mechanism of RTF was understood in terms of Zn vacancy in Mn-ZnO [19, 20]. There is a considerable report [15, 21] that attributed the coexisting impurity phases as the source for RTF in Mn-ZnO, other wise the single phase samples were suggested to be paramagnetic down to 5 K [22]. However, the role of secondary phase for RTF was ruled out by O.D. Jayakumar et al. [11]. They showed that RTF in Mn-ZnO nanomaterial arises due to the incorporation of $Mn^{2+}$ ions in the $Zn^{2+}$ ions matrix.

The most surprising feature is that magnetism in Mn-ZnO strongly depends on the material synthesis routes and conditions, irrespective of micron size (bulk) or nano size dimension of the particles [19, 22-24]. Hence, proper understanding of the structure and size-dependent properties are essential before applying DMS as the excellent material for next generation spintronic devices. In this case the route of material synthesis is not so much important, as much the reproducibility and stabilization of RTF is essential. Most of the DMS were synthesized following the bottom-up approach [25]. The properties of such DMS are not consistent to each other and also not free from structural defects, impurity phases and substrate effects. They need additional atmospheric environment which is not convenient for its accurate control during the material synthesis. On the other hand, top-down approaches such as mechanical milling and alloying have been recognized as the powerful tool for producing a variety of ZnO based DMS nanomaterials [26, 27]. This technique has effectively showed RTF in milled powder of Mn-ZnO, which needs further experimental works for its reproducibility and to understand structural changes and physical properties.

This paper deals with the synthesis of $Zn_{0.95}Mn_{0.05}O$ samples using mechanical milling and presents the structural and magnetic aspects of the samples. Mechanical allying was applied to confirm the reproducibility of the basic properties of mechanical routed DMS material. On viewing the room temperature ferromagnetism of the samples, attempt has been made to record the ferromagnetic to paramagnetic ordering temperature and such study is less explored in literature. The experimental results will be compared with reported works and will be understood in the frame work of existing models.

## 2. Experimental
## 2.1. Sample preparation:

Polycrystalline $Zn_{0.95}Mn_{0.05}O$ bulk sample was prepared using the standard solid-state reaction technique in argon atmosphere. The stoichiometric amounts of high purity

ZnO (99.99%, Alpha Aesar) and MnO (99.99% Alpha Aesar) powders were mixed thoroughly and fired around $900^0$C to $970^0$C for 24 hours. Then, the fired material was slowly cooled to room temperature and subsequently, made into powder by repeated grinding for 2 hours. The powder was made into pellets. Finally, the pellets were sintered at $970^0$C for 36 hours and furnace cooled. The crystalline phase of the final product was checked using X-ray diffraction spectrum. Being sure about the structural phase purity of $Zn_{0.95}Mn_{0.05}O$ bulk compound, Fritsch Planetary Mono Mill "Pulverisette 6" was used for mechanical milling. The milling was carried out in Argon atmosphere in an 80 ml agate vial with 10 mm agate balls. The ball to sample mass ratio was maintained to 7:1. For proper mixing, the milling process was stopped in every 4 hours interval. The milling was continued up to 60 hours. During this process, the vial was opened at 10, 20 and 35 hours to remove a small amount of material for checking the structural phase and grain size reduction. The samples are denoted as MH10, MH20, MH35 and MH60 for milling time 10, 20, 35 and 60 hours, respectively.

The nanocrystalline $Zn_{0.95}Mn_{0.05}O$ sample was also synthesized using mechanical alloying technique. Initially, the stoichiometric amounts of high purity ZnO and MnO powders were mixed using mortar and pestle for 2 hours. The mixed powder was mechanical milled for 72 hours with intermediate stopping and maintaining the identical conditions, as applied for mechanical milling of bulk sample. This alloyed sample with milling time 72 hours is denoted as MA72.

**2.2. Sample characterization and measurements**

The structural study of $Zn_{0.95}Mn_{0.05}O$ samples were carried out by recording the X-ray diffraction (XRD) spectrum at room temperature using X-ray diffractometer (model: X pert PANalytical). The XRD spectrum was recorded in the 2θ range 10 to $90^0$ with step size $0.01^0$ using Cu-K$_\alpha$ radiation (wavelength 1.54056 Å). The surface morphology of the samples was investigated by scanning electron microscopy (model: Hitachi S-3400N, Japan). The elemental compositions of the samples were carried out using Energy Dispersive analysis of X-ray (EDX) spectrum (Nortan System Six, Thermo electron corporation Instrument Super DRY II, USA). The elemental distributions of the samples were checked from the point and shoot microanalysis at ten different points and also using elemental mapping over a selected zone. The EDX spectrum of each sample was recorded at 10 various points. DC magnetization of the samples was studied as a function of magnetic field and temperature using vibrating sample magnetometer (Model: Lakeshore 7400), attached with low temperature cryostat and high temperature oven. The temperature dependence of magnetization was carried out at 2 kOe magnetic field by increasing the temperature from 100 K to 780 K and field dependence of magnetization was investigated within ± 15 kOe at room temperature.

## 3. Experimental results
### 3.1. Structural study

The elemental composition (Zn, Mn, O) of the samples is confirmed from the EDX spectrum, as shown in Fig. 1(a) for selected samples. The EDX spectra of the milled and alloyed samples are identical with respect to the bulk Mn-ZnO sample, without indicating any significant contamination (Si atom) from the agate balls and container. The elemental composition of the bulk and milled samples are close to that expected for $Zn_{0.95}Mn_{0.05}O$ compound, e.g. the atomic ratio of Zn, Mn and O is 50:47.53:2.47; 50:47.77:2.23; 50: 47.64:2.36; 50:47.60:2.40; 50:47.46:2.54 for bulk, MH20, MH35, MH60 and MA72 samples, respectively. Fig.1 (b) shows the XRD spectra of the selected $Zn_{0.95}Mn_{0.05}O$ (bulk, MH20, MH35, MH60 and MA72) samples. The XRD pattern of both bulk and mechanical (milled and alloyed) samples are identical (i.e., position, number and shape symmetry of the peaks). This indicates no significant additional phase, as reported earlier

[21, 23]. The crystalline cell parameters were obtained by matching the XRD peaks to single-phase hexagonal crystal structure with space group p63mc. The lattice parameters and unit cell volume of the $Zn_{0.95}Mn_{0.05}O$ samples are shown Fig. 2. The lattice parameters of bulk sample are $a$ = 3.247 Å and $c$ = 5.2083 Å. The general trend is that lattice parameter ($a$ and $b$) shows increasing trend with milling time and $c$ shows non-monotonic increase, which resulted in the increase of unit cell volume with milling time. The increasing effect of cell parameters is significant for the MA72 sample. The increase of lattice parameters suggests the incorporation of doped Mn atoms ($Mn^{2+}$ = 0.80 Å) in the Zn ($Zn^{2+}$ = 0.74 Å) sites of ZnO crystal structure [16, 28] and the process is enhanced by the milling induced effect (particle size reduction and lattice strain) [29].

The average grain size of the $Zn_{0.95}Mn_{0.05}O$ samples were estimated from the 6 to 7 prominent peaks of the XRD spectrum by using the Debye- Scherrer formula, $<d>_{Debye}$ = [0.089x180x$\lambda$/3.14x$\beta$x $cos\theta_C$] nm, where $2\theta_C$ is the position of peak center (in degrees), $\lambda$ is wavelength of X-ray radiation (1.54056 Å), $\beta$ is the full width at the half maximum of peak height (in degrees). The $2\theta_C$ and $\beta$ values were calculated by fitting the XRD peak profile to Lorentzian shape. The variation of grain size (using Debye formula) with milling time is shown in Fig. 2 (a). The grain size of $Zn_{0.95}Mn_{0.05}O$ sample slowly decreases with the increase of milling time, i.e., ($<d>_{Debye}$ ~ 25 nm for MH10, 21 nm for MH20, ~18 nm for MH35, ~14nm for MH60, and ~16nm for MA72 sample). The decrease of grain size may be consistent with the increased peak broadening of milled samples. The instrumental broadening and lattice strain are also expected to affect in peak broadening. Hence, Debye-Scherrer formula is not sufficient to accurately determine the grain size and lattice strain contribution. These aspects can be understood in a better way using Williamson-Hall plot, applied in other mechanical alloyed material [30, 31]. The Williamson-Hall equation is $\beta_{eff} \cos\theta_C = K\lambda/<d> + 2\varepsilon \sin\theta_C$, where K is Scherrer constant (0.89), $\varepsilon$ is the lattice strain and $\beta_{eff}^2 = \beta^2 - \beta_0^2$. In this case, $2\theta_C$ and $\beta$ were calculated by fitting the XRD peak profile to Voigt function and $\beta_0$ is the full width at half maximum of standard Silicon powder. The Voigt function is a convolution of a Lorentzian (L) function and a Gaussian (G) function. This corresponds to $\beta_L$ and $\beta_G$ for $\beta$ and subsequently, two values for $\beta_{eff}$. Finally, prominent XRD peaks (6 to 7 numbers) were used for the $\beta_{eff} \cos\theta_C$ vs. $2\sin\theta_C$ plot. The linear extrapolation to $\beta_{eff} \cos\theta_C$ axis gives $K\lambda/<d>$ (and hence, particle size $<d>$) and the slope gives the lattice strain ($\varepsilon$). The grain size ($<d>_L$ and ($<d>_G$ using $\beta_L$ and $\beta_G$ respectively) decreases with the increases of milling time (in Fig. 2 (a)). It is noted that $<d>_L$ is larger than $<d>_{Debye}$, where as the grain size $<d>_G$ becomes even smaller at higher milling time than $<d>_{Debye}$, although the nature of the variation with milling time is similar in all the cases. On the other hand, the initial increase of lattice strain ($\varepsilon_G$ using $\beta = \beta_G$) is followed by a slight decrease at higher milling time (Fig. 2(b)), where as $\varepsilon_L$ (using $\beta = \beta_L$) monotonically decreases with milling time. To extract meaningful information of XRD peak broadening due to simultaneous size and strain effects, it is assumed that Lorentzian (L) function is contributed solely due to grain size effects and Gaussian function (G) function is contributed due to lattice strain effect. Hence, we consider $<d>_L$ and $\varepsilon_G$ as the effective grain size and lattice strain, respectively, for the samples. The $<d>_L$ and $\varepsilon_G$ for MA72 sample are noted to be nearly 40 nm and 0.4, respectively and close to the value of MH60 sample.

Further, structural information (particle size, surface morphology and spatial distribution of elementals) is studied using SEM pictures of the samples. The SEM pictures (using 200 nm to 5 μm scale) of selected samples are shown in Fig. 3(a-d). The SEM picture of bulk sample (Fig.3 (a)) indicates the hexagonal crystal structure with size in the range ~1.84 μm to ~ 4.37 μm. The hexagonal shape of the crystallite domain transforms into more or less as spherical shape with smaller size for milled samples. The SEM picture of MH35 sample (Fig. 3(b)) shows the particle size in the range ~200 nm to ~ 435 nm. The particle size of the MH60 sample (Fig. 3(c)) is in the range ~100 nm to ~130 nm. The particle size of MA72 sample (Fig. 3(d)) varies in the range ~50 nm to ~ 80 nm. Hence, reduction of grain (single crystallite domain) size (from XRD data) is also confirmed from the SEM pictures. At the same time the estimated particle size (from SEM pictures) is larger than grain size from XRD data and agglomeration of particles or grains is indicated from the SEM pictures. This means SEM pictures suggest the size of multi-grained particles, which is expected to contribute more grain boundary effects. The line scan of EDX spectrum over a length 0 to 45 μm (Fig. 4(a-c)) suggests a reasonably good homogeneity of elements (Zn, Mn, O) distribution (within the experimental error due to varying surface roughness) in the milled samples. More interestingly, the mapping of elements, as shown in Fig. 4(d) for MH60 sample, indicated no appreciable clustering of Mn atoms in the ZnO matrix (mapping of O is not shown). This further suggests the appropriate doping of Mn in Zn sites in MH60 sample. We, now, forward to understand the ferromagnetism in Mn doped ZnO after having a sufficient amount of structural information of the samples.

**3.2. Magnetic study**

The room temperature (300 K) ferromagnetism in $Zn_{0.95}Mn_{0.05}O$ samples is confirmed from the magnetization (M) vs. applied field (H) curves (Fig. 5). First, we confirm the diamagnetism (negative slope in M(H) data) for bulk ZnO sample (inset of Fig. 5). Second, the ferromagnetism is confirmed at 300 K for Mn powder with spontaneous magnetization ($M_S$) ~ 0.089 emu/g (extracted from liner extrapolation of high field M(H) data to the M axis at H = 0), where as MnO powder shows expected paramagnetic response with field ($T_N$ ~ 120 K [24, 32]). The mixture of 5% MnO and 95% ZnO also showed paramagnetic behaviour (inset-a of Fig. 5). This indicates that the magnetism of the mixture is dominated by the paramagnetism of 5% MnO over the diamagnetism of 95% ZnO. Most interesting feature is that $Zn_{0.95}Mn_{0.05}O$ (Mn-ZnO) bulk sample exhibits (inset-a of Fig. 5) ferromagnetism with clear hysteresis loop and the shape is completely different from both Mn and MnO powders. The inset (b) of Fig. 5 shows a drastically different ferromagnetism of Mn-ZnO bulk sample in comparison with a typical ferromagnetic isotherm, characterized by a rapid increase of magnetization that saturates at higher magnetic field. The M(H) feature of bulk sample suggests some kind of induced ferromagnetism [30], as viewed (inset (b) of Fig. 5) from the extrapolation of (positive) high field M(H) data to negative M axis and negative high field M(H) data to positive M axis. The M(H) data at H ≥ +7 kOe is linearly fitted to the equation $M_L(H) = M_0 + \chi_0 H$ with $M_0$ = -0.006 emu/g and $\chi_0$ = 2.46x10$^{-6}$ emu/g/Oe for positive H values. The spontaneous magnetization ($M_S$ ~ 0.006 emu/g) of Mn-ZnO bulk sample is obtained from the polynomial fit of M(H) data at H < 7 kOe.

Now, we look at the room temperature ferromagnetism of milled samples. Fig. 6 shows that room temperature M(H) data of milled samples are different from the reference Mn-ZnO bulk and Mn powder samples. The M(H) isotherms at H ≥ 7 kOe is also linearly varied ($M_L(H) = M_0 + \chi_0 H$) for milled samples, where $M_0 = M_S$ and $\chi_0 H$ is the paramagnetic contribution superimposed on the ferromagnetic curve. The typical value of paramagnetic susceptibility ($\chi_0$) is ~ 1.79, 4.22, 4.21, 2.67 and 1.09 (in 10$^{-6}$ emu/g/Oe unit) for MH10, MH20, MH35, MH60 and MA72 samples, respectively. The effective ferromagnetic curve ($M_{eff}(H)$), as shown in the inset of Fig. 6, is obtained by subtracting the $\chi_0 H$ (H= 0 to 15 kOe) term from the observed M(H) data. The striking feature is that spontaneous magnetization (i.e., ferromagnetic ordering) has enhanced in milled samples. The obtained values of $M_S$ and other magnetic parameters from the M(H) loops, such as coercivity ($H_C$), remanent magnetization ($M_R$) and loop area (S) of the

$Zn_{0.95}Mn_{0.05}O$ samples are shown in Fig. 7. It is interesting to note that the magnetic parameters ($M_S$, $H_C$, $M_R$, S) all are showing increasing trend at lower milling time, where as the parameters are decreasing at higher milling time. The obtained coercivity in our samples is in the range of reported values [33, 34]. The magnetic parameters of MA72 sample is significantly lower in comparison with the MH60 sample, suggesting more soft ferromagnetic nature of MA72 sample.

The ferromagnetic ordering of $Zn_{0.95}Mn_{0.05}O$ samples below and above of room temperature is investigated from the temperature (100 K to 800 K) dependence of dc magnetization curves at 2 kOe (Fig. 8). DC magnetization of the samples decreased continuously when the temperature increases up to the ferromagnetic to paramagnetic transition temperature ($T_C$), as indicated by arrow for each sample. The $T_C$ of Mn-ZnO bulk sample ~ 500 K is higher in comparison with the reported value ~380 K [35] and may be comparable to the predicted value above 425 K [10]. The typical M(T) behaviour with a slight up curvature at lower temperature indicates the coexistence of paramagnetic contribution with ferromagnetism [36]. The M(T) shape is completely changed for milled samples (down curvature), indicating the appearance of dominant ferromagnetic order. It is highly remarkable that $T_C$ of the material is enhanced (~ 595 K, 605 K, 620 K and 640 K for MH10, MH20, MH35 and MH60 samples, respectively) in the mechanical milled nanoparticle samples. The above experimental results (enhanced ferromagnetic moment and ordering temperature) are supportive to the earlier reports that ferromagnetism in doped DMS is strongly affected by disorder, which increases $T_C$ of the material [37, 38]. The enhancement of $T_C$ on the particle size reduction by mechanical route is different in comparison with a few preliminary reports that suggested the decrease of $T_C$ in chemical routed nanoparticle sample [35]. It is worthy to mention that surface magnetism (also magnetic ordering) of mechanical milled samples and chemical routes samples can not be compared [39], as attempted in some report [21]. The ferromagnetic order above room temperature is also reproduced in the present work for MA72 sample with $T_C$ ~ 615 K.

## 4. Discussions

Based on our experimental results of mechanical milled samples, the XRD spectra do not show any detectable additional phase at room temperature. The additional phase is well pronounced in the XRD spectrum of the same compound in those reports, where RTF was attributed due to impurity phase: like $Zn_xMn_{3-x}O_4$ [15, 21, 23, 40]. It may be mentioned that $Zn_xMn_{3-x}O_4$ like spinel phase is paramagnetic at room temperature [41]. In those reports the starting materials were $MnO_2$ and ZnO. Similar starting materials were used in previously reported mechanical milled Mn-ZnO compounds [33, 34, 42, 43] and different features were noted at room temperature. In our case, the mechanical alloyed samples (started with raw ZnO and MnO powders) also reproduced the structural phase purity and room temperature magnetism (RTF), as observed for milled samples. Hence, the effect of additional phases is not applied for the present samples. The present study shows that mechanical milling and alloying technique is much more effective in producing the nanomaterials of DMS in the sense that it excludes the complexity and phase destabilization related to the high temperature annealing effects [15, 21, 23]. The role of impurity during milling process in producing the RTF is also ruled out, as the contamination (Si) is minor (below 1 atomic %) and non-magnetic. Although some recent report [44] suggested the appearance of magnetic signal in ZnO crystal due to Ar (1 to $2 \times 10^{17}$ $Ar/cm^2$) implantation, but the magnetic moment in the present samples is more than 100 times stronger than that observed due to high dose Ar implantation. The RTF in Mn-ZnO samples, prepared under different atmospheres [24, 33, 34], suggests the inconclusive role of any specific atmosphere for exhibiting RTF. The effect of Ar atmosphere during mechanical milling is believed to be not significant for the observed RTF and its existence up to 640 K in our samples. The minor role of the impurity phase and atmospheric condition, if effective at al, can be treated as the additional disorder of nanoparticles. Hence, the basic origin of RTF in transition metal doped ZnO is something else, which is not clear till date.

Now, we proceed to understand the observed RTF in the presented samples. The Mn-ZnO bulk sample exhibited an unconventional M(H,T) feature, which is viewed as consisting of ferromagnetic order superimposed with a strong paramagnetic contribution. The result is supportive to the theoretical prediction of room temperature paramagnetism and ferromagnetism for p–type Mn-ZnO [3, 10, 13, 20]. The room temperature M(H) of bulk sample is resemble to the modulated magnetic order appeared due to the diffusion of (high magnetic moment) Fe atoms (from $Fe_2O_3$ canted ferromagnet) into the sites of (low magnetic moment) Cr atoms (from $Cr_2O_3$ paramagnet at 300 K) [30] of rhombohedral (hexagonal) lattice structure. This cationic ($Fe^{3+}$ and $Cr^{3+}$) diffusion process can be compared to the diffusion of $Mn^{2+}$ ions into $Zn^{2+}$ sites [11, 24]. For the sake of argument, if there is no diffusion among Mn and Zn before and after forming the bulk compound, then only paramagnetic response was expected. But, the M(H) shape of bulk sample is completely different from the mixture of 95%ZnO and 5% MnO. This experimental data definitely indicate the introduction (alloying) of Mn atoms into Zn sites and confirmed in other reports [11]. The alloy formation is checked from XRD data and a good homogeneity of elemental distribution is seen from the EDX spectrum. Such alloy formation results in the modification of magnetic interactions among (magnetic) Mn atoms. The spontaneous magnetic moment (~ 0.006 emu/g ~ $1.7 \times 10^{-3}$ $\mu_B$/Mn atom) and moment at 15 kOe (including paramagnetic component) ~ 0.032 emu/g ~ $9.0 \times 10^{-3}$ $\mu_B$/Mn atom in bulk Mn-ZnO sample is much smaller than the expected 5 $\mu_B$/Mn atom at 0 K for high spin $Mn^{2+}$ sates [44]. This means Mn atoms are either in low spin states [32] or moment of the Mn atoms is quenched in the non-favourable (diamagnetic) environment of Zn. The best example of the effect of environment on magnetic ordering can be exchange bias system, where ferromagnetic (Co) core is surrounded by antiferromagnetic (CoO) shell [45]. The enhancement of magnetic moment ($M_S$ ~ 0.0385 $\mu_B$/Mn atom and ~ 0.055 $\mu_B$/Mn atom at 15 kOe for MH35 sample) and ordering temperature up to 640 K in mechanical milled/alloyed nanoparticle samples can be explained using the core-shell model of nano size grains. For the present samples, a less ferromagnetic core ($Zn_{0.95}Mn_{0.05}O$) is surrounded by more ferromagnetic shell ($Zn_{0.95}Mn_{0.05}O$). The notable decrease of room temperature magnetic parameters ($M_S$, $H_C$, $M_R$ and loop area) at higher milling time represents the increase of shell disorder. However, the magnetic parameters were noted to be larger for milled samples in comparison with un-milled bulk sample. The lattice expansion in nanoparticle samples, arises due to shell disorder [29] and strain induced anisotropy [39], is favourable for the enhanced ferromagnetic interactions [20, 32, 37, 46] among the (core-shell) interfacial Mn atoms. This increases the effective number of ferromagnetic Mn atoms in the shell and the moment per Mn atom in the nanoparticle samples. It may be mentioned that EDX spectrum is taken in the micron scale and the local distribution of cations is averaged out in that spectrum. Hence, the cation vacancies at the atomic scale may not be reflected in the EDX spectrum. The disorder (cation vacancy and other lattice defects) in shell structure can promote the required number of hole carriers, as predicted for carrier mediated ferromagnetism in Mn doped ZnO [3, 10, 13]. It is unlike in our samples that (95%) Zn atoms diffuses into the surface of (5%) $MnO_2$ particles and double exchange type mechanism is responsible for RTF in Mn doped ZnO [34, 40, 42], because double exchange mechanism has shown the maximum $T_C$ in perovskite compound up to 370 K [46], where as the $T_C$ in our samples is up to 640 K. J.M.D. Coey [47] proposed that lattice defects can be favourable for the enhanced long range ferromagnetic ordering of magnetic moments. From the present study, we believe that the (core-shell) interfacial disorder of nanoparticles strongly affects the diffusion of Mn atoms into the lattices of Zn. The disorders in shell might be able to produce the impurity band that can mediate a long range ferromagnetic ordering of the Mn moments [19]. On the other hand, ferromagnetism with $T_C$ up to 620 K was noted in double perovskite ($Sr_2BB^/O_6$; B= Fe, Cr; $B^/$ = Mo, Re) materials [48, 49], where super exchange type interactions between B and $B^/$ ions via O anion (affected by band hybridization and electron doping) determine the ferromagnetic ordering temperature, inspite of increasing disorder in the lattice structure. Some attempts have been made [3, 20, 50-52] to understand the ferromagnetic ordering in transition metal doped ZnO taking into account the increasing density of states at the Fermi level, spin polarization of electrons at the conduction band and carrier mediated ferromagnetic exchange interactions. However, further experiments and theoretical studies are needed to establish such approaches in confirmed manner.

## 4. Conclusion

The mechanical milling and alloying is capable of producing nanoparticles of $Zn_{0.95}Mn_{0.05}O$ (DMS) without showing any significant additional phase. Ferromagnetism at and above room temperature is confirmed in $Zn_{0.95}Mn_{0.05}O$ sample and supporting many theoretical prediction. The coexisting paramagnetic contributions are accounting the quenching of magnetic moment per Mn atom in $Zn_{0.95}Mn_{0.05}O$. The secondary phase is not playing any major role in showing the ferromagnetism in the present samples. The ferromagnetic ordering is attributed to the diffusion of $Mn^{2+}$ ions into the Zn sites of ZnO lattice structure and the diffusion process is enhanced by the milling induced effects (particle size reduction, grain boundary disorder and mechanical strain). To search for the basic origin of RTF, it would be interesting if the mechanisms for high temperature ferromagnetism are generalized for different magnetic oxides (DMS, double perovskites). It is anticipated that such challenging problem could be taken care by many theoretical models and experimental works in future.

**Acknowledgment:** The authors thank to CIF, Pondicherry University for providing the experimental facilities. RNB and AS also thank to UGC for financial support [F.NO. 33-5/2007 (SR)].

**References:**

[1]     G. A. Prinz, Science **282**, 1660 (1998).

[2]     J. K. Furdyna, J. Appl. Phys. **64**, R29 (1998).

[3]     T. Dietl, H. Ohno, F. Matsukura, J. Cibert, and D. Ferrand, Science **287**, 1019 (2000).

[4]     K. Ueda, H. Tabata, and T. Kawai, Appl. Phys .Lett. **79**, 988 (2001).

[5]     V. A. L Roy, A. B. Djurisic, H. Liu and X. X. Zhang, Appl. Phys. Lett. **84**, 756 (2004).

[6]     M. Hung, S. Mao, H. Feick, H. Yan, Y. Wu, H. Kind, E. Weher, R. Russo, and P. Yang, Science **292**, 1897 (2001).

[7]     X.Wen, Y. Fang, Q. Pang C. Yang, J. Wang, W. Ge, K.S. Wong and S. Yang, J. Phys. Chem. B  **109**, 15303 (2005).

[8]     C. R. Gorla, N. W. Emanetoglu, S. Liang , W. E. Mayo, Y. Lu, M. Wraback, and H. Shen,  J. Appl. Phys. **85**, 2595 (1999).

[9]     X. D. Bai, P. X. Gao, Z. L. Wang and E. C. Wang, App.Phys.Lett. **82**, 4806 (2003).

[10]    P. Sharma, A. Gupta, K. V. Rao, F. J. Owens, R. Sharma, R. Ahuja, J. M. O. Guillen, B. Johansson, and G. A. Gehring, Nat. Mater. **2**, 673 (2003).

[11]    O.D. Jayakumar, H.G. Salunke, R. M. Kadam, M. Mohapatra, G. Yaswant and S. K. Kulshreshtha, Nanotechnology **17,** 1278 (2006).

[12]   M. Ivill, S. J. Pearton, Y. W. Heo, J. Kelly, A. F. Hebard, and D. P. Norton, J. Appl. Phys. **101**, 123909 (2007).

[13]   T. M. Souza, I. C. da Cunha Lima, and M. A. Boselli, Appl. Phys. Lett. **92**, 152511 (2008).

[14]   K. Sato and H. Katayama-Yoshida, Physics E (Amsterdam) **10**, 251 (2001).

[15]   D. C. Kundaliya, S. B. Ogale, S. E. Lofland, S. D. Dhar, C. J. Metting, S. R. Shinde, Z. Ma, B. Varughese, K. V. Ramanujachari, L. Salamance-Riba, and T. Venkatesan, Nat. Mater. **3**, 709 (2004).

[16]   J. Luo, J. K. Liang, Q. L. Liu, F. S. Liu, Y. Zhang, B. J. Sun, and G. H. Rao, J. Appl. Phys. **97**, 086106 (2005).

[17]   N. A. Spaldin, Phys. Rev. B **69**, 125201 (2004).

[18]   G. Lawes, A. S. Risbud, A. P. Ramirez, and R. Seshadri, Phys. Rev. B **71**, 045201 (2005).

[19]   Q. Xu, H. Schmidt, L. Hartmann, H. Hochmuth, M. Lorenz, A.Setzer, P. Esquinazi, C. Meinecke, and M.Grundmann, Appl. Phys. Lett. **91**, 092503 (2007).

[20]   D. Iuşan, B. Sanyal, and O. Eriksson, Phys. Rev. B **74**, 235208 (2006)

[21]   J. Blasco, F. Bartolomé, Luis M. Garcı́a and Joaquı́n Garcı́a, J. Mater. Chem. **16**, 2282 (2006).

[22]   S. J. Han, T. H. Jang, Y. B. Kim, B. G. Park, J. H. Park, and Y. H. Jeonga, Appl. Phys. Lett. **83**, 920 (2003).

[23]   S. K. Mandal, A. K. Das, T. K. Nath, D. Karmakar, B. Satpati, J. Appl. Phys. **100**, 104315 (2006).

[24]   W. Chen, L. F. Zhao, Y. Q. Wang, J. H. Miao, S. Liu, Z. C. Xia, and S. L. Yuan, Appl. Phys. Lett. **87**, 042507 (2005).

[25]   Z. R. Dai, Z. W. Pan, and Z. L. Wang, Adv. Funct. Mater. **13**, 9 (2003).

[26]   L. C. Damonte, L. A. M. Zelis, B. M. Soucase and M. A. H. Fenollosa.,Powder Technol. **148**, 15 (2004).

[27]   A. M. Glushenkov, H. Z. Zhang, J. Zou, G. Q. Lu, and Y. Chen, Nanotechnology **18**, 175604 (2007).

[28]     R. Viswanathan, S. Sapra, S. Sen Gupta, B. Satpati, P.V. Satyam, B. N. Dev and D. D. Sarma, J. Phys. Chem. B **108**, 6303 (2004).

[29]     R. N. Bhowmik and R. Ranganathan, and R. Nagarajan, Phys. Rev. B **73**, 144413 (2006).

[30]     R. N. Bhowmik, N. Murty and S. Srinadhu, PMC Physics B **1,** 20(2008).

[31]     S. Vives, E. Gaffet and C. Meunier, Mater.Sci.Eng. A **366**, 229 (2004).

[32]     M. A. Morales, R. Skomski, S. Fritz, G. Shelburne, J. E. Shield, Ming Yin, Stephen O'Brien, and D. L. Leslie-Pelecky, Phys. Rev. B **75,** 134423 (2007).

[33]     H. J. Blythe, R. M. Ibrahim, G. A. Gehring, J. R. Neal, A. M. Fox, J. Magn. Magn. Mater. **283,** 117 (2004).

[34]     K. Tanaka, K. Fukui, S. Murai, and K. Fujita, Appl. Phys. Lettr. **89**, 052501 (2006).

[35]     V.K. Sharma, R. Xalxo and G.D. Varma, Cryst. Res. Technol. **42**, 34 (2007).

[36]     N.A. Theodoropoulou, A. F. Hebard, D. P. Nortan, J. D. Budai, L. A. Boatner, J. S. Lee, Z. G. Khim, Y. D. Park, M. E. Over berg, S. J. Pearton and R. G. Wilson, Solid State Electronics **47**, 2231 (2003).

[37]     M. Berciu and R. N. Bhatt, Phys. Rev. Lett. **87**, 107203 (2001).

[38]     D. Iuşan, B. Sanyal, and O. Eriksson, Phys. Rev. B **74**, 235208 (2006).

[39]     R. N. Bhowmik, R. Ranganathan, R. Nagarajan, B. Ghosh, and S. Kumar, Phys. Rev. B **72**, 094405 (2005).

[40]     M. A. García, M. L. Ruiz-González, A. Quesada, J. L. Costa-Krämer, J. F. Fernández, S. J. Khatib, A. Wennberg, A. C. Caballero, M. S. Martín-González, M. Villegas, F. Briones, J. M. González-Calbet and A. Hernando, Phys. Rev. Lett. **94**, 217206 (2005).

[41]     S. Asbrink, A. Waskowska, L. Gerward, J. S. Olsen and E. Talik, Phys. Rev. B **60**, 12651 (1999).

[42]     A. Quesada, M.A. García, P. Crespo, A. Hernando, J. Magn. Magn. Mater. **304,** 75 (2006).

[43]     F. Bartolomé, J. Blasco, L.M. García, J. García, S. Jiménez, A. Lozano, J. Magn. Magn. Mater. **316,** e195 (2007).

[44]     R. P. Borges, R. C. da Silva, S. Magalhaes, M. M. Cruz and M. Godinho, J. Phys.: Condens. Matter **19,** 476207 (2007).

[45]     V. Skumryev, S. Stoyanov, Y. Zhang, G. Hadjipanayis, D. Givord, and J. Nogues Nature **423,** 850 (2003).

[46]     J. Mira, J. Rivas, F. Rivadulla, C. Vázquez-Vázquez, and M. A. López-Quintela,  Phys. Rev. B **60** 2998 (1999)

[47]     J. M. D. Coey,  Solid State Sci. **7,** 660 (2005).

[48]     J. M. De Teresa, D. Serrate1, C. Ritter, J. Blasco, M. R. Ibarra, L. Morellon, and W. Tokarz,  Phys. Rev. B **71**, 092408 (2005).

[49]     E. K. Hemery, G. V. M. Williams and H. J. Trodahl  Phys. Rev. B **74**, 054423 (2006).

[50]     Navarro,  J. Fontcuberta,  M. Izquierdo, J. Avila, and M. C. Asensio, Phys. Rev. B **69**, 115101 (2004).

[51]     X. Zhang, W. Fan, S. Li, and J. Xia, Appl. Phys. Lett. **91**, 223103 (2007).

[52]     A. J. Behan, A. Mokhtari, H. J. Blythe, D. Score, X-H. Xu, J. R. Neal, A. M. Fox, and G. A. Gehring, Phys. Rev. Lett. **100,** 047206 (2008).

Figure Captions

Fig.1. (a) EDX spectra and (b) XRD spectra for selected samples.

Fig. 2(a) Variation of lattice parameters (*a* & *c*) and (b) unit cell volume (V) with milling hours for Zn0.95Mn0.05O samples. The dotted lines guide the general trend of *a* and *V*.

Fig.3 (Colour online) Grain size (<d>) (a) Strain (e) of the samples with milling hours, estimated by Williamson-Hall plot. The full width at half maximum of XRD peak was calculated using Lorentzian function (Debye method) and voigt (Lorentzian + Gaussian components) function.

Fig. 4 (a-c). (Colour online) Elemental scan [O(red), Mn (green), Zn (blue-K line and magenta-L line) as indicated In (c) for MA72 sample] over the length 0 to 45 μm. (d) elemental mapping for MH60 sample.

Fig. 5 (Colour online) Field dependence of Magnetization for Mn and MnO powders (main panel), Mn-ZnO bulk sample and mixture (inset-a), and Mn-ZnO bulk sample for high field range (inset-b) and line is the extrapolation of high field M(H) data.

Fig. 6 (Colour online) Field dependence of magnetization for different samples (main panel) and inset shows the M(H) data (0 to 15 kOe range) after subtracting linear component from high field (> 7 kOe) data.

Fig.7 Variation of Magnetic parameters, i.e., (a) spontaneous magnetization ($M_S$), (b) coercivity ($H_C$), (c) remanent magnetization ($M_R$), and (d) hysteris loop area, with milling time and calculated M(H) data at 300 K.

Fig.8 Temperature dependence of zero field cooled magnetization at 2 kOe for different $Zn_{0.95}Mn_{0.05}O$ samples. The arrows mark the $T_C$ of respective sample.

Fig. 1

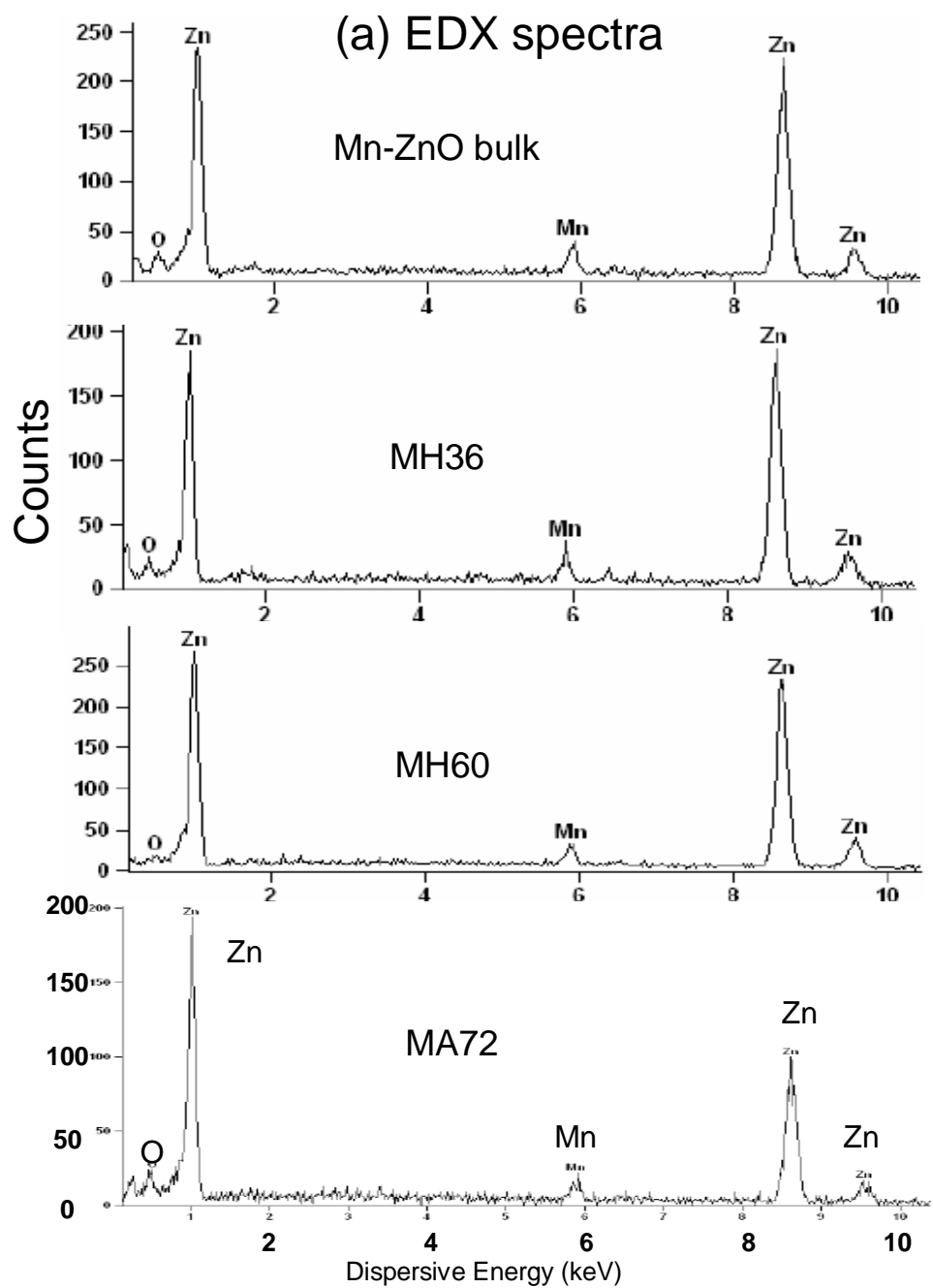
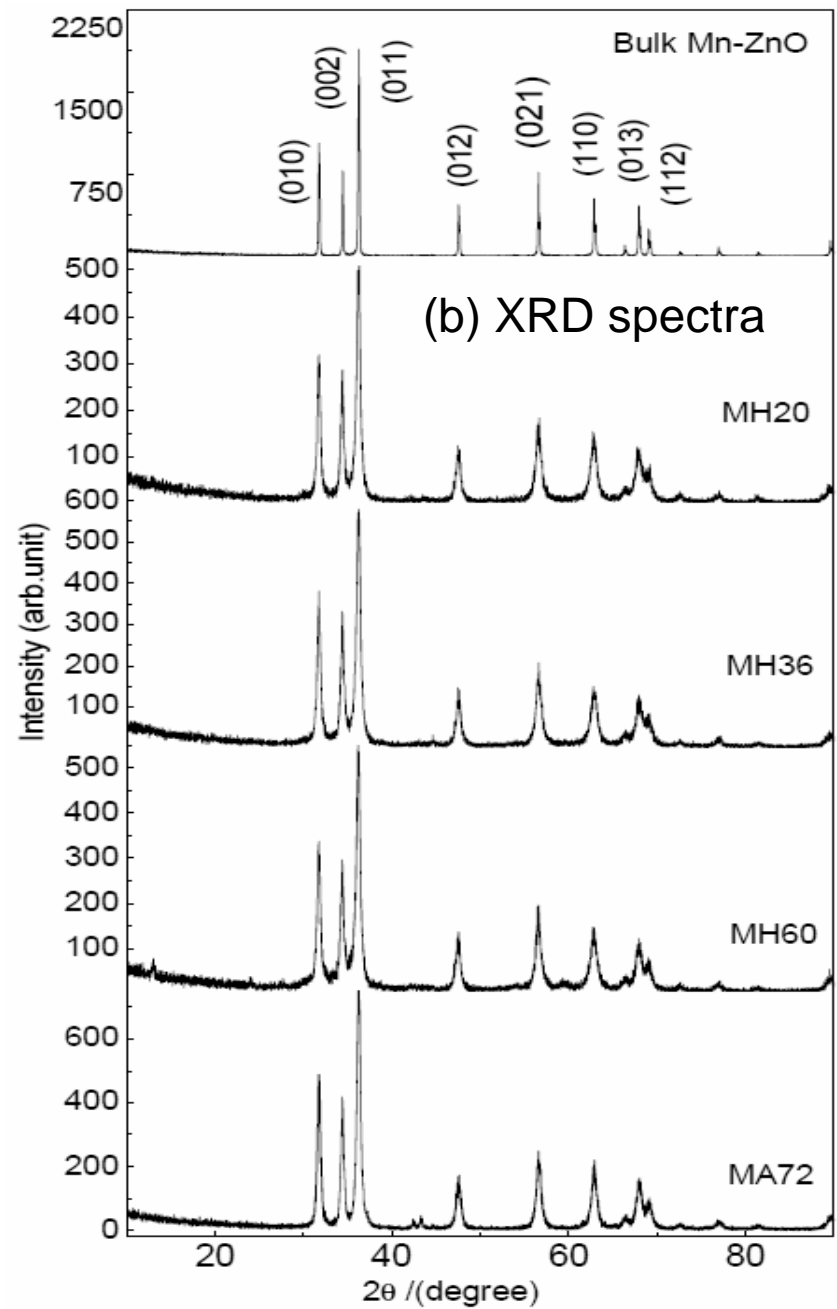

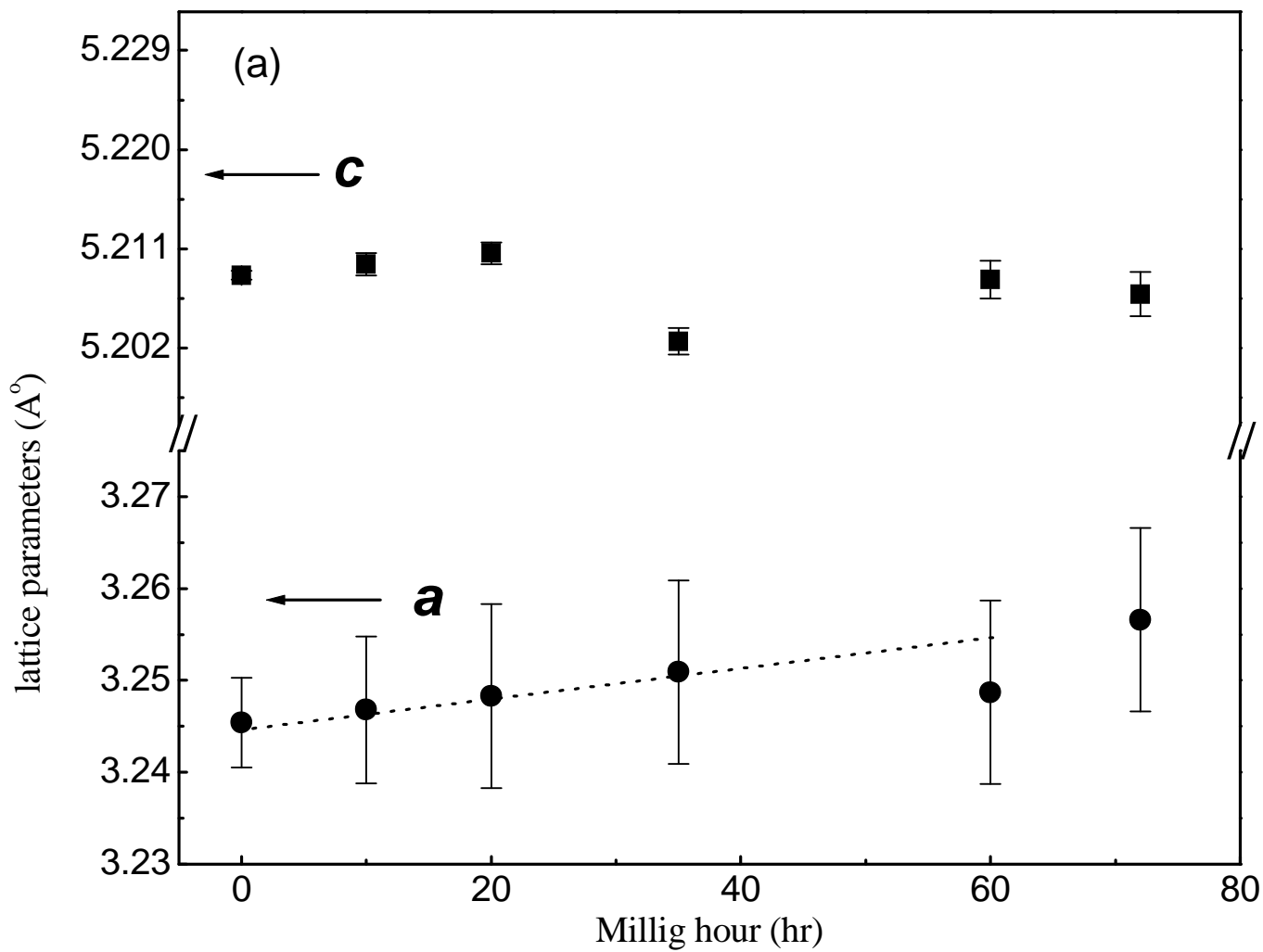
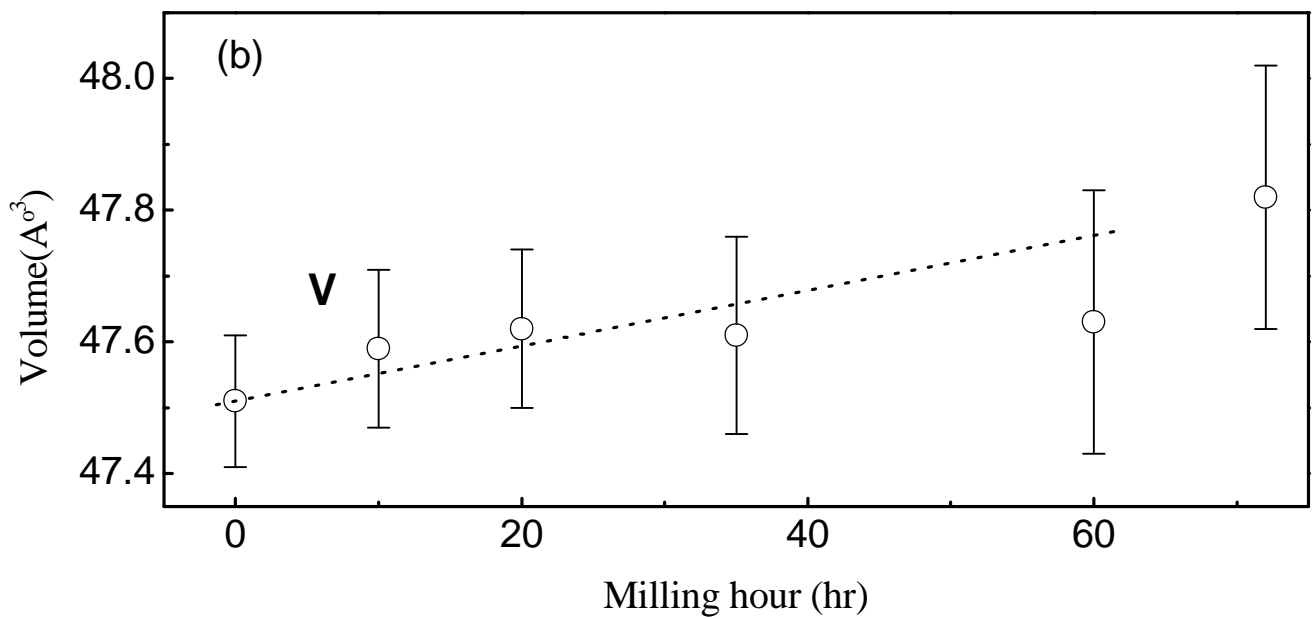

Fig. 2(a) Variation of lattice parameters (a & c) and (b) unit cell volume (V) with milling hours for $Zn_{0.95}Mn_{0.05}O$ samples. The dotted lines guide the general trend of *a* and *V*.

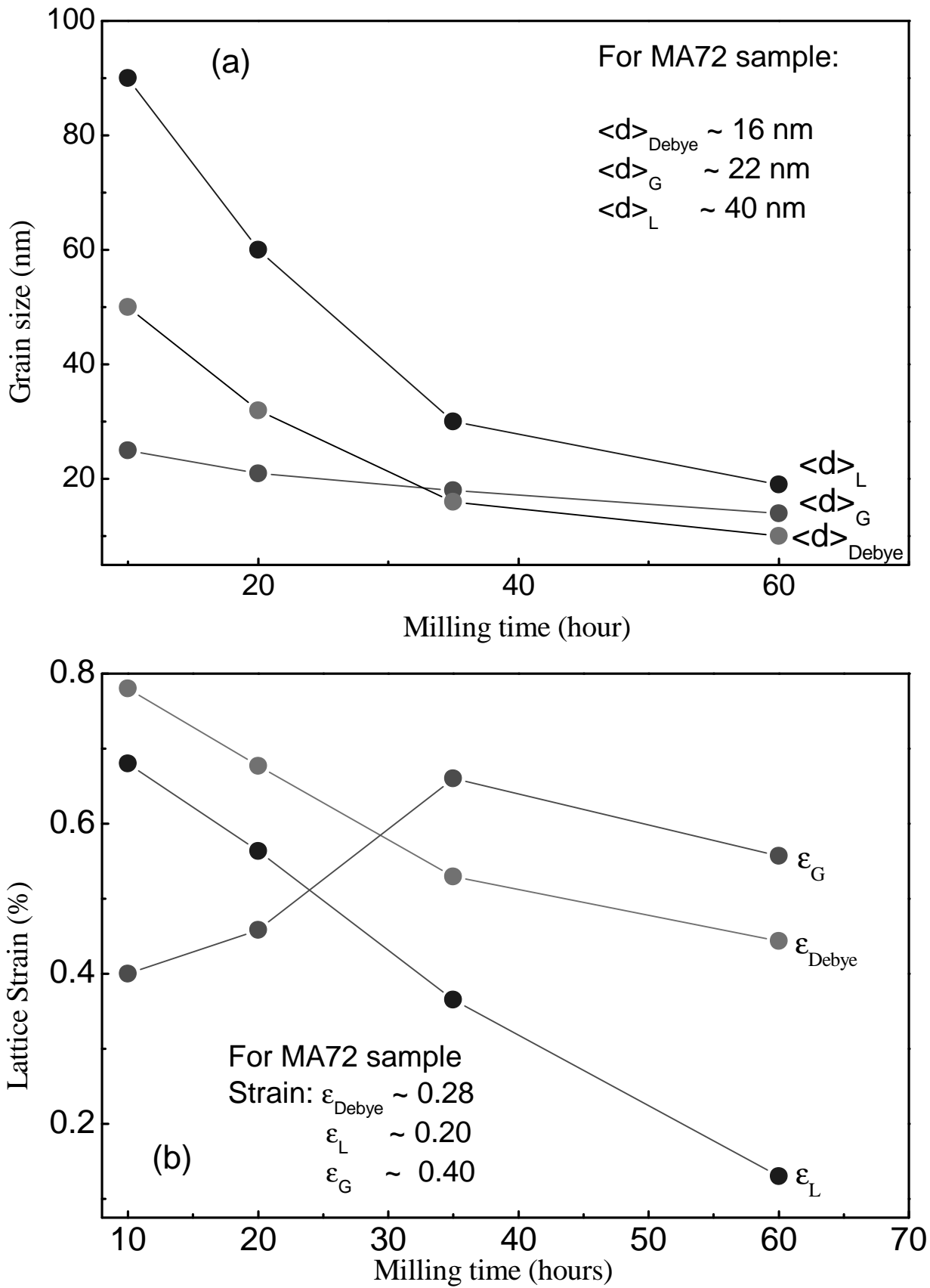

Fig.3 (Colour online) Grain size (<d>) (a) Strain (ε) of the samples with milling hours, estimated by Williamson-Hall plot. The full width at half maximum of XRD peak was calculated using Lorentzian function (Debye method) and voigt (Lorentzian + Gaussian components) function.

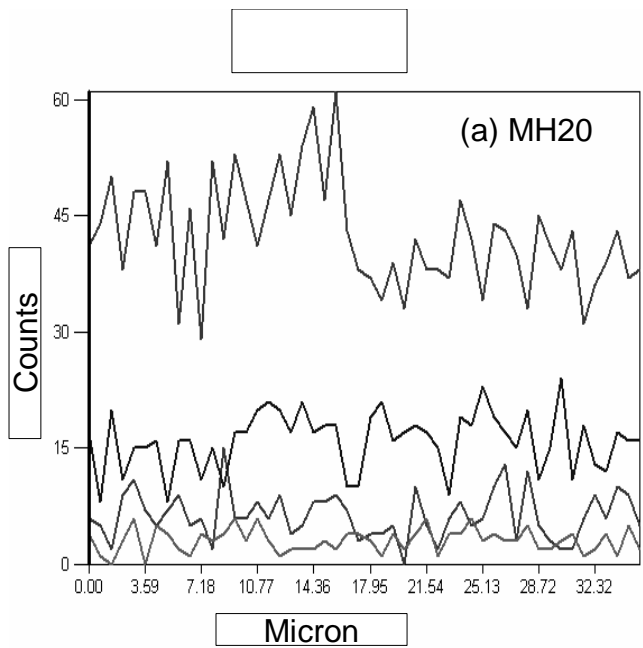
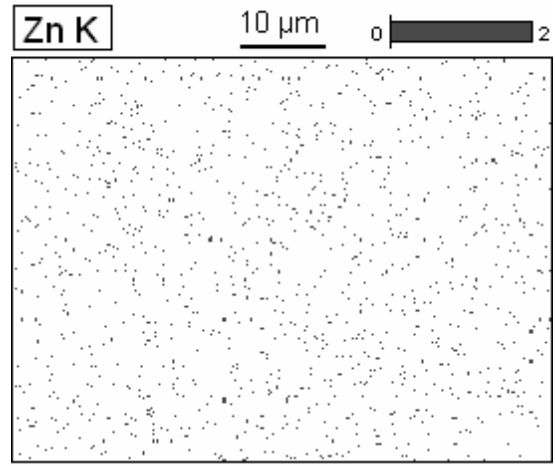
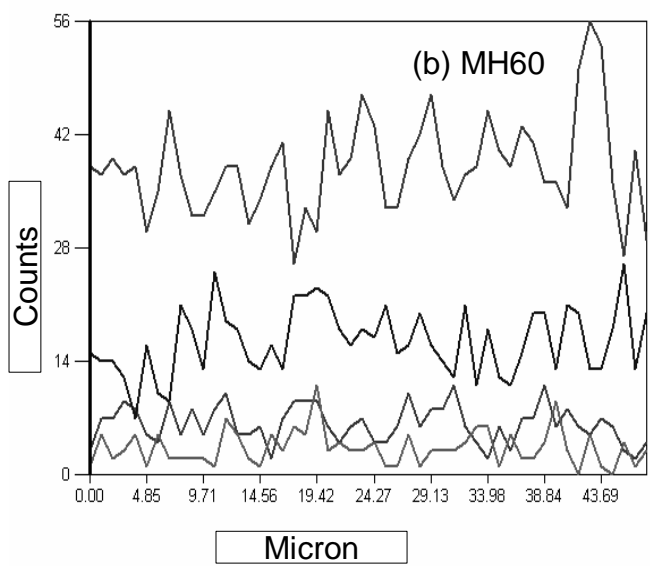
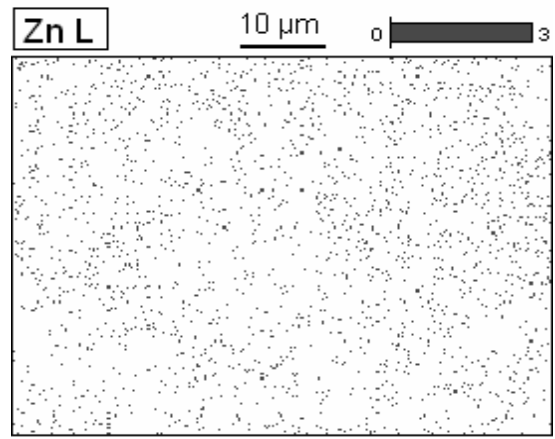
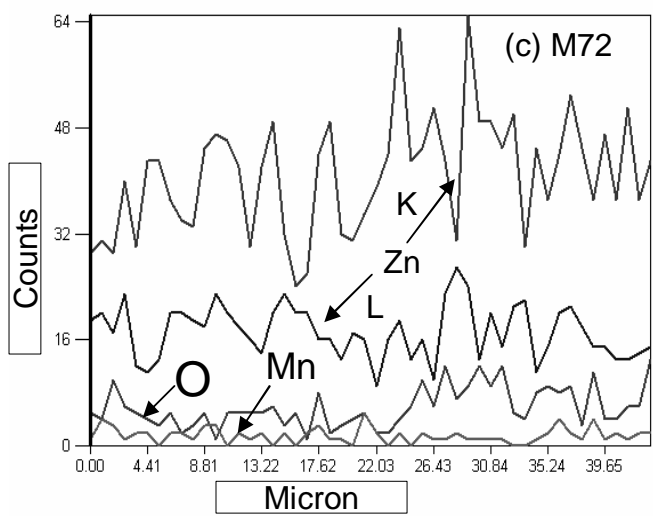
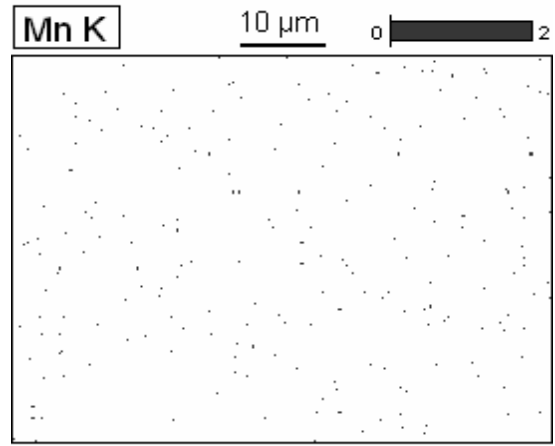

Fig. 4 (a-c). (Colour online) Elemental scan [O(red), Mn (green), Zn (blue-K line and magenta-L line) as indicated In (c) for MA72 sample] over the length 0 to 45 μm. (d) elemental mapping for MH60 sample.

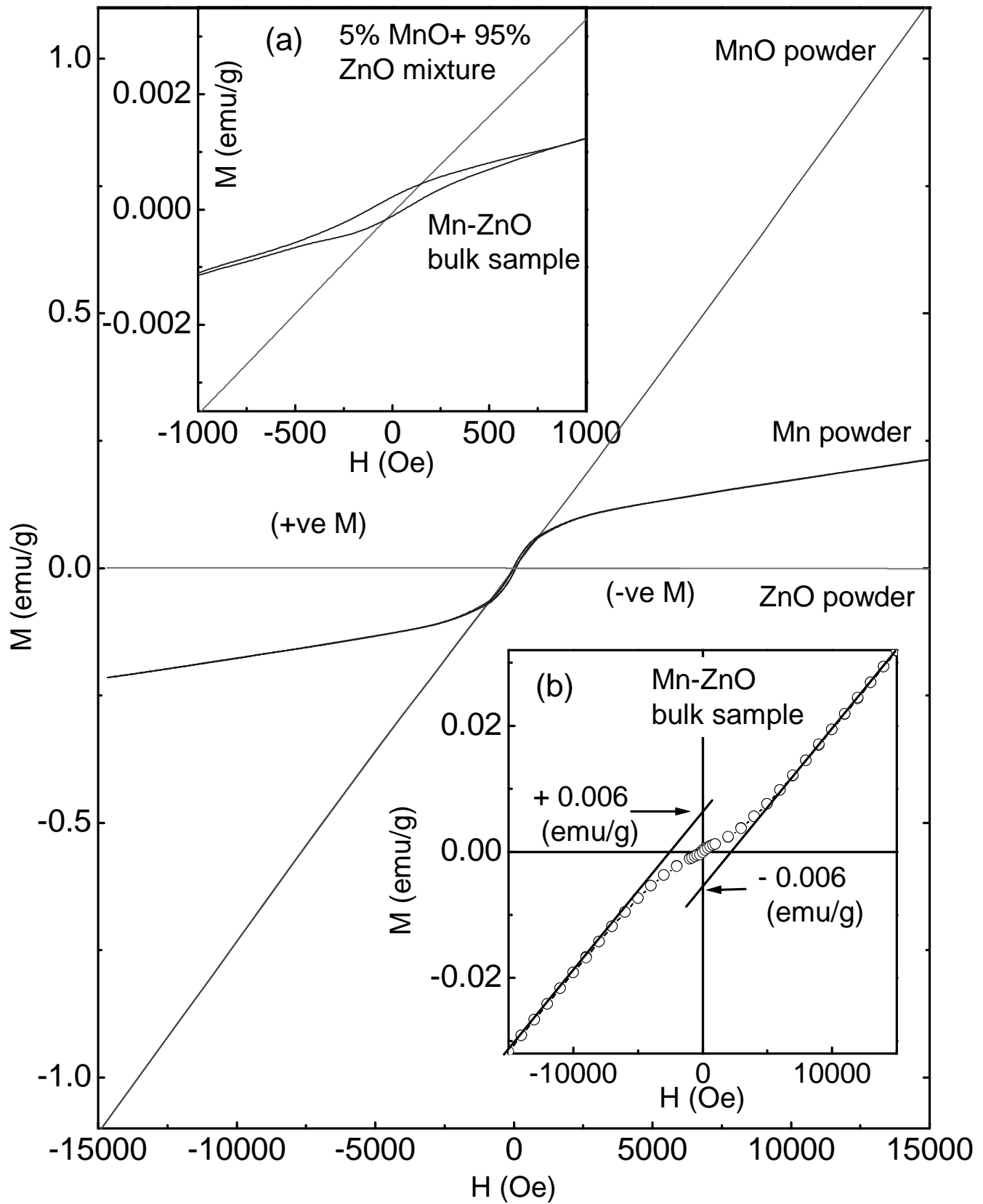

Fig. 5 (Colour online) Field dependence of Magnetization for Mn and MnO powders (main panel), Mn-ZnO bulk sample and mixture (inset-a), and Mn-ZnO bulk sample for high field range (inset-b) and line is the extrapolation of high field M(H) data.

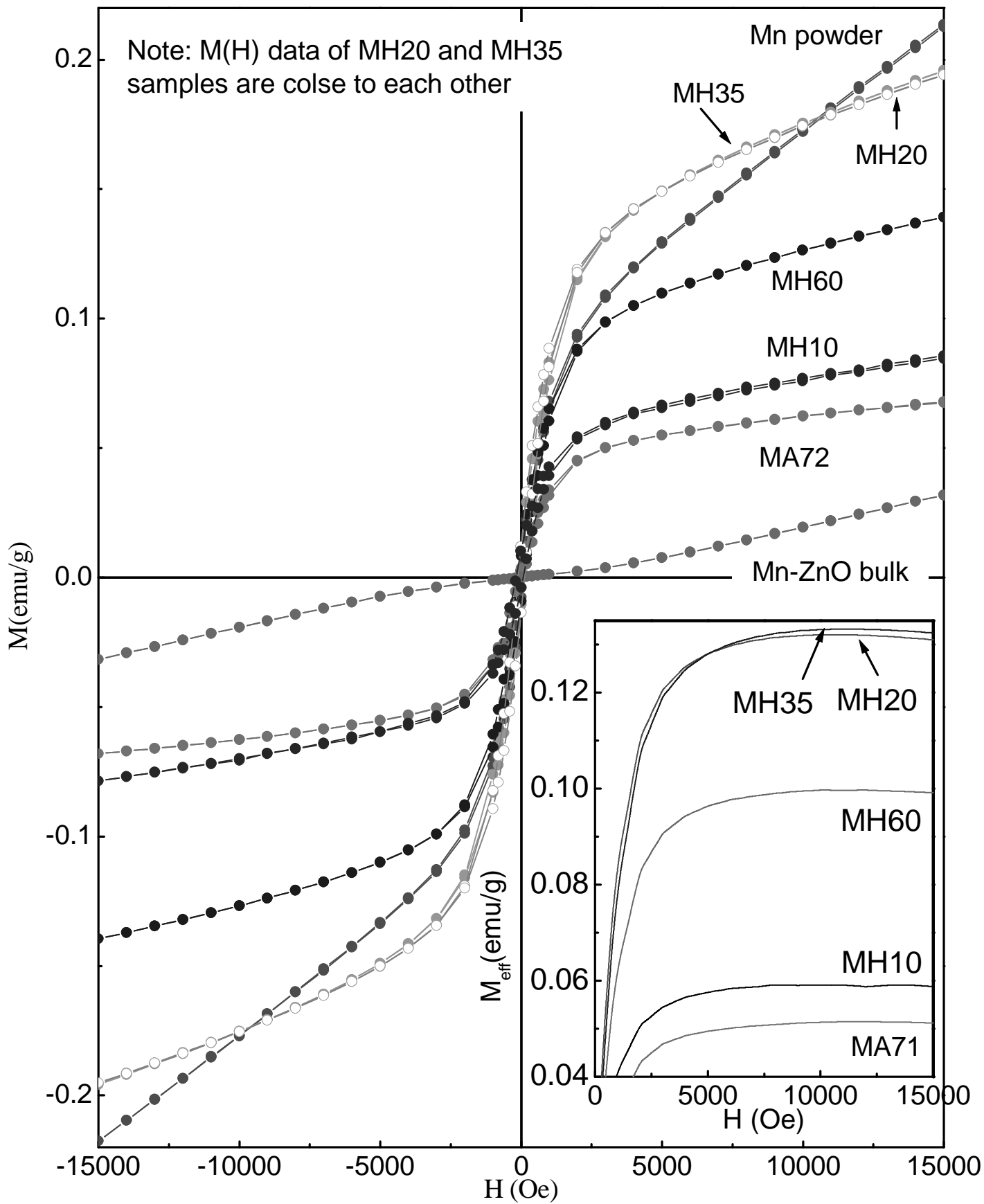

Fig. 6 (Colour online) Field dependence of magnetization for different samples (main panel) and inset shows the M(H) data (0 to 15 kOe range) after subtracting linear component from high field (> 7 kOe) data.

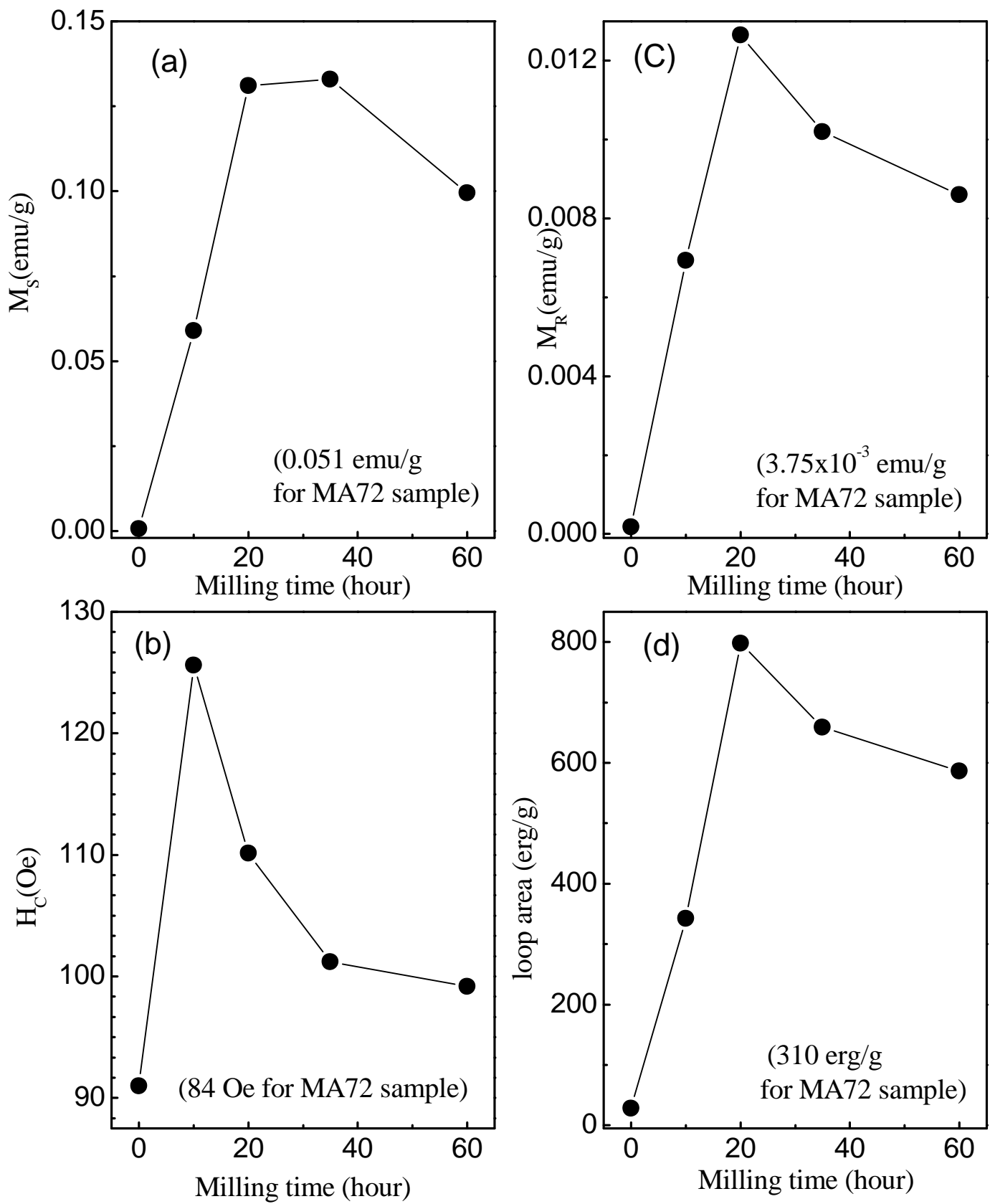

Fig.7 Variation of Magnetic parameters, i.e., (a) spontaneous magnetization ($M_S$), (b) coercivity ($H_C$), (c) remanent magnetization ($M_R$), and (d) hysteris loop area, with milling time and calculated M(H) data at 300 K.

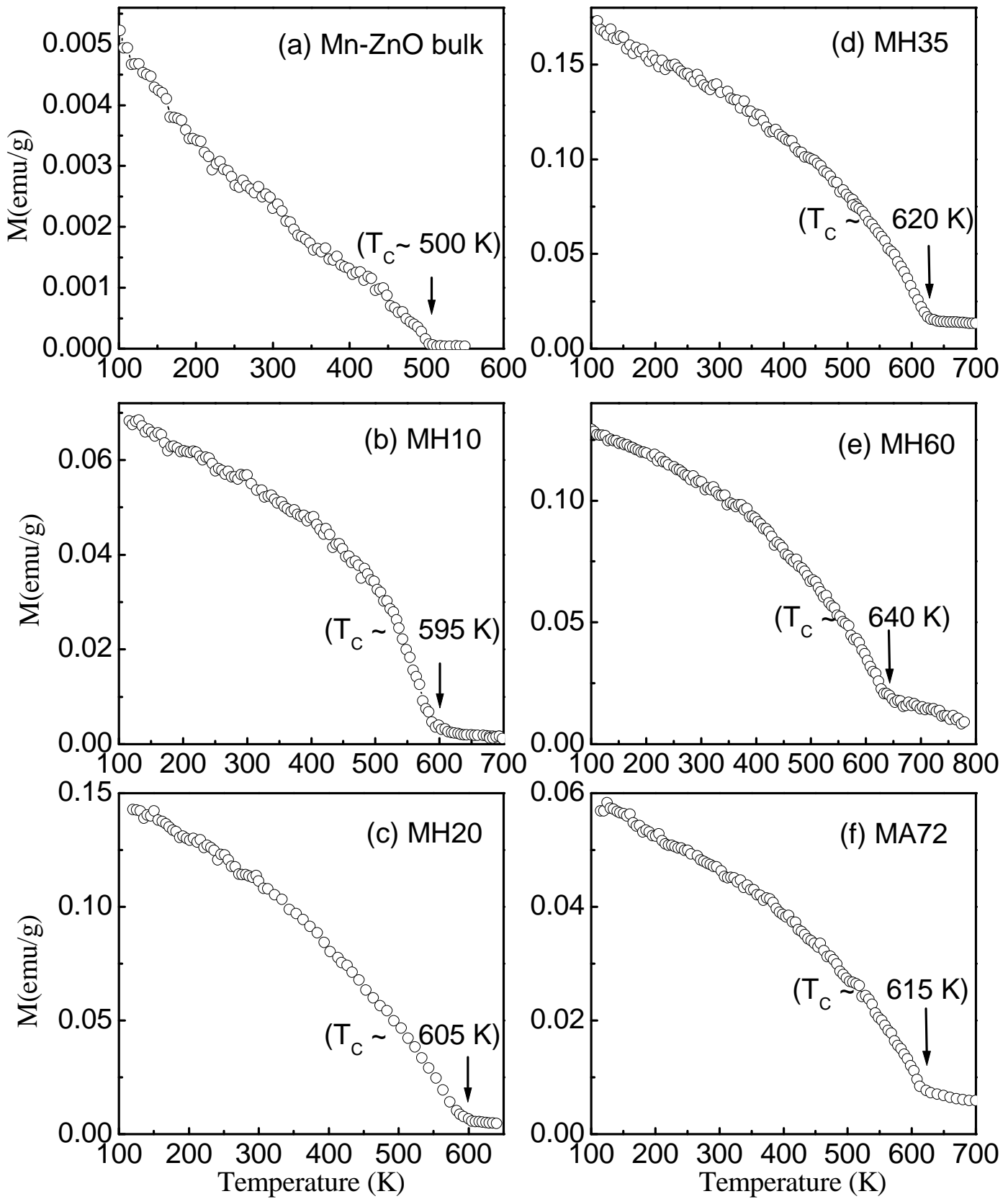

Fig.8 Temperature dependence of zero field cooled magnetization at 2 kOe for different $Zn_{0.95}Mn_{0.05}O$ samples. The arrows mark the $T_C$ of respective sample.